\documentclass[11pt,a4paper]{article}
\usepackage{lmodern}
\usepackage[english]{babel}
\usepackage[margin=2cm]{geometry}
\usepackage{pdflscape} 
\usepackage{amsmath}
\usepackage{amssymb}
\usepackage{graphicx}
\usepackage[colorlinks=true, allcolors=blue]{hyperref}
\usepackage{xr}
\usepackage{float}
\usepackage{siunitx}
\usepackage{multirow}
\usepackage{subcaption}
\usepackage{longtable}
\usepackage[version=4]{mhchem} 
\usepackage{authblk}
\usepackage{orcidlink}
\usepackage{rotating}
\usepackage[normalem]{ulem}

\title{
Screening novel cathode materials from the Energy-GNoME database using MACE machine learning force field and DFT
}

\author[1,*]{Nada Alghamdi\ \orcidlink{0000-0001-5638-3483}}
\author[1,*]{Paolo de Angelis\ \orcidlink{0000-0003-1866-2988}}
\author[1,2]{Pietro Asinari\ \orcidlink{0000-0003-1814-3846}}
\author[1,2,*]{Eliodoro Chiavazzo\ \orcidlink{0000-0001-6165-7434}}

\affil[1]{Department of Energy, Politecnico di Torino, Corso Duca Degli Abruzzi, 10129 Torino, Italy}
\affil[2]{Istituto Nazionale di Ricerca Metrologica, Strada delle Cacce 91, 10135 Torino, Italy}

\date{}
\begin{document}
\maketitle
\vspace{-2.5em}
\begin{center}
*Corresponding authors: \texttt{nada.alghamdi@polito.it}, \texttt{paolo.deangelis@polito.it},\texttt{eliodoro.chiavazzo@polito.it}
\end{center}


\begin{abstract}
The development of new battery materials, particularly novel cathode chemistries, is essential for enabling next generation energy storage technologies.
In this work, we employ a multi-fidelity screening protocol combining the Energy-GNoME confident criteria, foundational MACE machine-learning force fields (MLFF), and physically motivated heuristic filters to identify novel intercalation cathodes for post-lithium batteries, namely: Na-, K-, Mg-, and Ca-ion batteries.
Foundational MACE models are used to efficiently asses dynamical stability, thermodynamical stability, average voltage, and theoretical specific energy, enabling a rapid screening of candidates. 
For the most promising cathodes, voltage predictions are refined using DFT+U calculations.
%
This work delivers three key outcomes: i) establishing and validating a robust high-throughput screening approach for cathode materials with foundational MLFF models;
ii) suggestions for cathode candidates for the development of next-generation of batteries;
iii) a fair comparison between the MACE predictions and the readily available figures of merit reported in the Energy-GNoME database on the examined materials.
\end{abstract}

\section{Introduction}

Driven by the transition to electric mobility and the expansion of stationary grid storage, the global demand for high-performance and compact energy storage is increasing rapidly \cite{Fichtner2022Rechargeable,Amici2022roadmap,koech_lithium-ion_2024}. 
Recent consolidated data from the International Energy Agency (IEA) indicate that global battery demand surpassed 1 TWh per year in 2024~\cite{IEA2025CriticalMinerals}.
Projections from the stated policies scenario (STEPS) and similar studies indicate higher estimates~\cite{koech_lithium-ion_2024, IEA2025CriticalMinerals, IEA2023WEO}.
A rapid increase is expected, with demand tripling over the next few years to reach approximately 6~TWh per year by 2040~\cite{IEA2023WEO,koech_lithium-ion_2024}.
This growth is accompanied by increasingly stringent requirements for higher specific energy, lower cost, longer cycle life, sustainable and secure raw-material supply, and improved safety.
While incremental improvements in cathode chemistries and manufacturing technologies continue to enhance performance, Lithium-ion batteries (LIBs) still rely heavily on critical and geographically concentrated materials, which increases cost volatility, supply-chain risk, and sustainability concerns as production scales up~\cite{IEA2025CriticalMinerals,ngoy_lithium-ion_2025}.
These constraints, together with technical challenges related to long-term stability, fast charging, and second-life management are driving electrochemical and materials scientists to explore alternatives to conventional LIB frameworks and investigate new chemistries for developing post-lithium batteries for next-generation energy storage systems~\cite{Fichtner2022Rechargeable,ngoy_lithium-ion_2025}.

In response to these challenges, other alkali-metal  charge carriers, most notably sodium (Na) and potassium (K), are attracting increasing attention as sustainable alternatives to lithium due to their natural abundance, low cost, and uniform geographic distribution~\cite{Kubota2018towards}.
They can offer higher ionic conductivity compared to Li-ions due to smaller ionic radius when transporting in various solvents~\cite{Kubota2018towards}.
While Na-ion batteries typically exhibit slightly lower specific energy compared to LIBs, finding new materials with higher voltage and specific capacity can make Na-ion batteries advantageous to implement over conventional LIBs especially for high-power applications~\cite{Kubota2018towards, Goikolea2020Na}.
Additionally, K-ion systems, promise specific energy almost identical to Li-ion batteries~\cite{Kubota2018towards}. 
Moreover, because Na- and K-ion batteries use similar materials to their Li-ion counterparts, they can readily leverage the existing manufacturing pipelines and infrastructure~\cite{Kubota2018towards}.
On the other hand, multivalent ion systems, like magnesium (Mg) and calcium (Ca), are gaining more traction as promising energy storage systems thanks to their low cost, abundance, and improved safety~\cite{Chen2021Emerging}.
They exhibit low reactivity to ambient environment, compatible with non-flammable aqueous electrolytes, and the metal anodes for Mg and Ca show dendrite-free plating~\cite{Chen2021Emerging, Liang2020Multivalent}.
Finding electrode materials, however, that exhibit both good operating voltages and allow for rapid and reversible ion intercalation remains an open challenge~\cite{Chen2021Emerging}.
Therefore, in search for next-generation of batteries, finding new materials, especially new cathodes, is paramount~\cite{Kubota2018towards, Chen2021Emerging}.
Predictive computational tools and material screening of novel cathode materials stands out as an indispensable approach for accelerating this discovery process~\cite{Fichtner2022Rechargeable}.

Materials discovery has traditionally utilized a forward-design approach, in which starting from a set of structures their properties are evaluated for specific applications~\cite{Curtarolo2013HighThroughput, cheng2025aidrivenmaterialsdesignminireview}.
The datasets to be screened may consist of experimental databases, which are explored to identify previously overlooked applications of existing materials, or hypothetical, computationally generated materials to enable the rational design of experimental new materials.
Crystal structure prediction methods are used for the creation of new hypothetical material candidates by determining the crystal structure starting from the desired number of atoms and atomic types~\cite{oganov2018crystal, davies2016computational, Oganov2019StructurePrediction}.
The exploration of crystalline inorganic materials has been transformed by artificial intelligence (AI) \cite{cheng2025aidrivenmaterialsdesignminireview}.
Whether within the forward design scheme by constructing candidate structures using AI approaches~\cite{merchant2023scaling, Chen2025Crystal} and faster methods for evaluating their properties \cite{mace-foundations},
or through the rise of inverse-design methods aimed at predicting structures that meet targeted application specific requirements~\cite{DeBreuck2025GenerativeReview}.

A prominent example of an AI-driven construction of hypothetical crystals is the graph networks for materials exploration (GNoME)~\cite{merchant2023scaling} from Google DeepMind that used an iterative active learning pipeline with graph neural networks (GNNs) to discover stable crystals~\cite{merchant2023scaling}. 
While GNoME have claimed to identify hundreds of thousands of stable crystalline structures~\cite{merchant2023scaling}, a key step that still remains is the assignment to those material of possible technological applications as well as synthesis pathways~\cite{Cheetham2024Artificial}.
To address the former challenge on practical applications, the Energy-GNoME initiative~\cite{DEANGELIS2025100605,de_angelis_energy-gnome_2024} has leveraged the concept of cross-domain data bias \cite{Trezza2025Classification} to create a database of materials specifically designed for energy applications. 
So far, it has identified 20,454 promising cathode materials suitable for lithium and post-lithium (e.g., Na, Mg, K, Ca) battery systems~\cite{DEANGELIS2025100605} (with confidence above 50\%).
At this stage, however, the performance and feasibility of these candidates rely primarily on predictions from machine-learning (ML) models.
Although the protocol proposed by 
Energy-GNoME~\cite{DEANGELIS2025100605, de_angelis_energy-gnome_2024} seeks to mitigate common ML limitations, such as out-of-distribution effects, the predictions remain inherently data-driven and may be less reliable for emerging battery chemistries (e.g. Ca-ion cathodes) where the available training data are comparatively sparse.
To progress beyond this pre-screening, a more physics-grounded verification is essential. 
Although density functional theory (DFT)~\cite{hohenberg1964, kohn1965} remains the gold standard for such validation, its high computational cost presents a major bottleneck for rapid screening dozens of thousands of materials or large sized systems.

To this end, we leverage on MACE foundational models~\cite{Batatia2025foundation}, which are machine learning force field (MLFF) based on the message passing atomic cluster expansion~\cite{Batatia2022mace} trained on large diverse chemistries.
Foundational MLFFs offer a powerful tool for high-throughput screening \cite{JACOBS2025101214}.
Several foundational MACE models have emerged based on different datasets.
The earliest models, such as MACE-MP-0a and MACE-MP-0b, that were trained exclusively on the Materials Project trajectory (MPtrj) dataset \cite{deng2023chgnet} showed wide-range applicability \cite{Batatia2025foundation}.
More recent models have incorporated out-of-equilibrium structures and higher-level functionals. 
These include the MACE-OMAT-0 (MACE-OMAT), trained on the large OMat24 dataset~\cite{barrosoluque2024openmaterials2024omat24} containing over 100 million data points, and the MACE-MATPES-r2SCAN-0 (MACE-r2scan), trained on the MATPES-r2SCAN dataset \cite{kaplan2025foundationalpotentialenergysurface}.
By leveraging foundational MLFF training on extensive and diverse datasets, these models facilitate rapid property prediction, enabling the efficient identification and filtering of unsuitable candidates~\cite{JACOBS2025101214}.

Therefore, in this work, we use MACE foundational models for screening the cathode candidates in the Energy-GNoME database~\cite{DEANGELIS2025100605,de_angelis_energy-gnome_2024}.
First, we rationalize the validity and accuracy of our approach against experimentally established and characterized cathodes.
Following that, we continue to screen the Energy-GNoME cathode candidates and subsequently refine the average voltage prediction of the most promising candidates using DFT.
This work builds on and contributes to the growing body of research on AI-driven materials discovery, which has become increasingly important with the advent of generative AI. 
While such models can generate large numbers of hypothetical crystals, physically grounded analyses are still required~\cite{DeBreuck2025GenerativeReview,zeni2025generative,Mikkel2026Continued}.

\section{Results and discussion}
\subsection{Overview of the screening protocol}
\begin{figure*}[!ht]
   \centering
   \includegraphics[width=\linewidth]{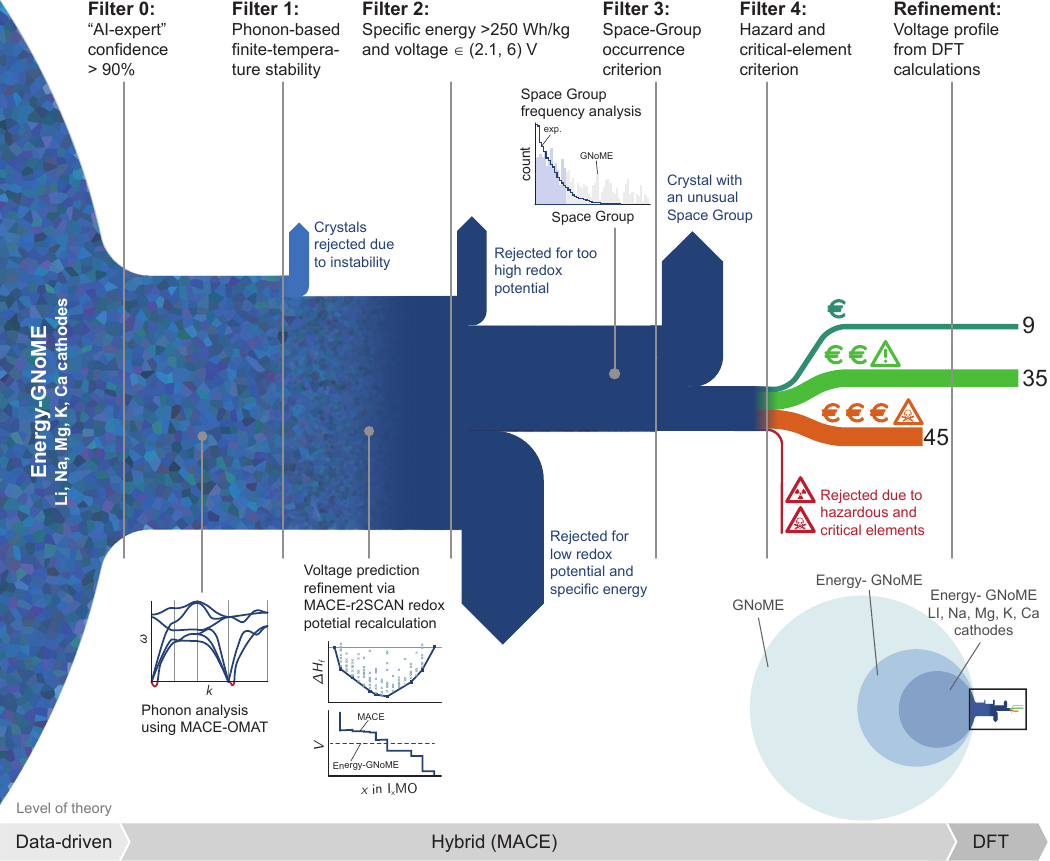}
   \caption{Overview of the screening protocol. The flow diagram shows the progressive reduction of the candidate pool across the different screening stages, with flow thickness representing relative pool size. Below is indicated the level of theory used at each step, while the bottom-right zoomed-out schematic provides an approximate visualization of the candidate pool cardinality relative to the initial GNoME database.}
   \label{fig:workflow}
\end{figure*}

This section provides an overview of the screening protocol implemented in this study.
Our multi-level screening protocol is depicted in Fig.~\ref{fig:workflow}, the Energy-GNoME candidates are progressively refined through a sequence of increasingly stringent filters that substantially reduce the size of the candidate pool. 

In the first step, cathodes containing Cs, Al, Rb, and Y are excluded, together with all candidates exhibiting an average AI-expert confidence below 90\% (\textit{Filter 0} in Fig.~\ref{fig:workflow}), leaving only 615 crystals.
This confidence score can be interpreted as the probability that a given crystal represents a possible cathode material.
The screening then proceeds by employing the foundational MLFF model MACE as a surrogate model to replace the substantially more expensive DFT calculations. 
Thanks to its physics-informed formulation and training on diverse chemical environments, MACE enables rapid yet reliable evaluation of key thermodynamic and vibrational properties at scale, as demonstrated in Section. \ref{Validation of computational methods}.
Using MACE, we apply two physics-based screening filters:
\begin{itemize}
    \item \textit{Filter 1} assesses dynamical stability by computing phonon dispersions from MACE force calculations on relaxed crystal structures.
    Thereby excluding dynamically unstable candidates.
    \item \textit{Filter 2} evaluates the electrochemical performance by computing equilibrium voltage profiles for each remaining candidate.
    This allows us to improve upon the Energy-GNoME voltage estimates and exclude cathodes with average voltages below 2.1 V or above 6 V, as well as materials with theoretical specific energies lower than 250 Wh/kg.
\end{itemize}

Following these physics-based criteria, we apply two additional heuristic filters motivated by experimental materials science considerations.
\begin{itemize}
    \item \textit{Filter 3} leverages the statistical distribution of space groups observed in experimentally reported inorganic crystals. Candidates exhibiting symmetries outside the 24 most frequently occurring space groups (accounting for two thirds of known inorganic crystal structures)~\cite{urusov2009frequency} are excluded, as such symmetries are less likely to correspond to experimentally synthesizable materials~\cite{Cheetham2024Artificial, HICKS2021110450, ECKERT2024112988}.
    \item \textit{Filter 4} focuses on the elemental composition of the remaining candidates. Each material is classified into three categories according to chemical hazard, raw-material cost, and elemental availability~\cite{BORAH2020OnBattery}. Candidates containing toxic or radioactive elements are discarded at this stage (red branch in Fig.~\ref{fig:workflow}). 
\end{itemize}
The candidates composed exclusively of abundant and low-cost elements (dark green branch in Fig.~\ref{fig:workflow}) and candidates containing at least one marginally costly or marginally toxic element (green branch in Fig.~\ref{fig:workflow}) move forward to the final refinement stage, in which the equilibrium voltage profiles are recomputed using density functional theory with Hubbard corrections (DFT+U), providing a more accurate assessment of the electrochemical performance. 
At the end of the complete screening workflow, only few tens of candidates survive, corresponding to an overall reduction of nearly four orders of magnitude with respect to the initial materials space (GNoME database) and yielding a tractable set of cathode materials suitable for experimental investigation.

\subsection{Validation of computational methods}
\label{Validation of computational methods}
\subsubsection{Voltage profile and specific energy}
\begin{figure}[!h]
    \centering
    \includegraphics[width=0.475\linewidth]{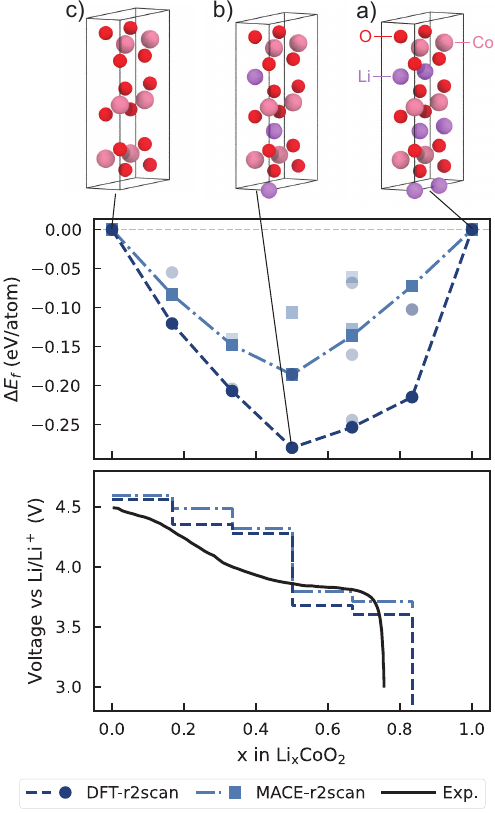}
    \caption{
        The thermodynamic phase stability of \ce{LiCoO2} during Li de-intercalation (LIB charging) is illustrated through convex-hull constructions (top) and the corresponding voltage profile (bottom) computed at the two levels of theory: DFT in dark blue and MLIP, i.e. MACE, in blue. 
        Representative compositions along the de-intercalation pathway are highlighted as (a) fully intercalated \ce{LiCoO2}, (b) half de-intercalated \ce{Li_{0.5}CoO2}, and (c) fully de-intercalated \ce{CoO2}.
        The experimental voltage in solid black line profile is provided for comparison, using data taken from Ref.~\cite{HU202361}.
    }
    \label{fig:convex_hull_ave_V_LiCoO2}
\centering
\end{figure}

We start by validating our method for the average voltage and specific energy, two key metrics for battery performance, on six experimentally realizable cathode materials. 
This includes three lithium-based cathodes: the first commercial cathode material \ce{LiCoO2} \cite{Lyu2021OverviewLiCO2}, the safer and more stable \ce{LiFePO4} \cite{ZHANG20112962}, and the less conventional \ce{Li2MnO3} \cite{Li2MnO3}\cite{Rana2014Structural}.
In addition, to examine generalizability to other chemistries, we also selected three experimentally characterized cathodes based on different charge-carriers: sodium-ion \ce{NaCoO2} \cite{reddy2015high}, potassium-ion \ce{KVPO4F} \cite{K-ion-cathode}, and magnesium-ion \ce{MgV2O4} \cite{Hu2020High}.
The charging of the battery, corresponding to ion de-intercalation, was simulated using an iterative semi-brute force method.
In contrast to the brute force method that explores all possible ion arrangements and hence scale exponentially with the number of sites $\mathcal{O}(2^{N_{sites}})$.
In our approach, each subsequent de-intercalation step (ion removal) is initiated from the most stable structure found in the immediately preceding step.
The resulting scaling is quadratic in the number of sites $\mathcal{O}(N_{sites}^2)$.
For the supercells considered in this study, which contain up to $N_{sites} = 30$ ions, a full brute-force search corresponds to over $10^{8}$ possible configurations whereas the semi-brute force method reduces the required configurations to approximately 900.
The convex hull based on the formation energies (see Sec. \ref{method:Convex_hull_and_voltage_profile}) was constructed.
The structures lying on the hull represent the thermodynamically stable intermediate phases.
The combined voltage profile and convex hull for \ce{LiCoO2} is shown in Fig. \ref{fig:convex_hull_ave_V_LiCoO2}.
Interestingly, the MACE-r2scan closely follows the experimental and DFT curves.
The other cathode materials maintain good performance, with varying levels of agreement (Fig. S1).
Notably, the MACE-r2scan calculations were completed in minutes on a single NVIDIA A100 GPU, whereas the analogous DFT-r2scan calculation required several days.
While careful distribution of the de-intercalation steps can reduce the DFT time, a full voltage profile for even a small system still demands a few hours with optimal parallelization.

\begin{figure}[!htb]
    \centering
    \includegraphics[width=0.5\linewidth]{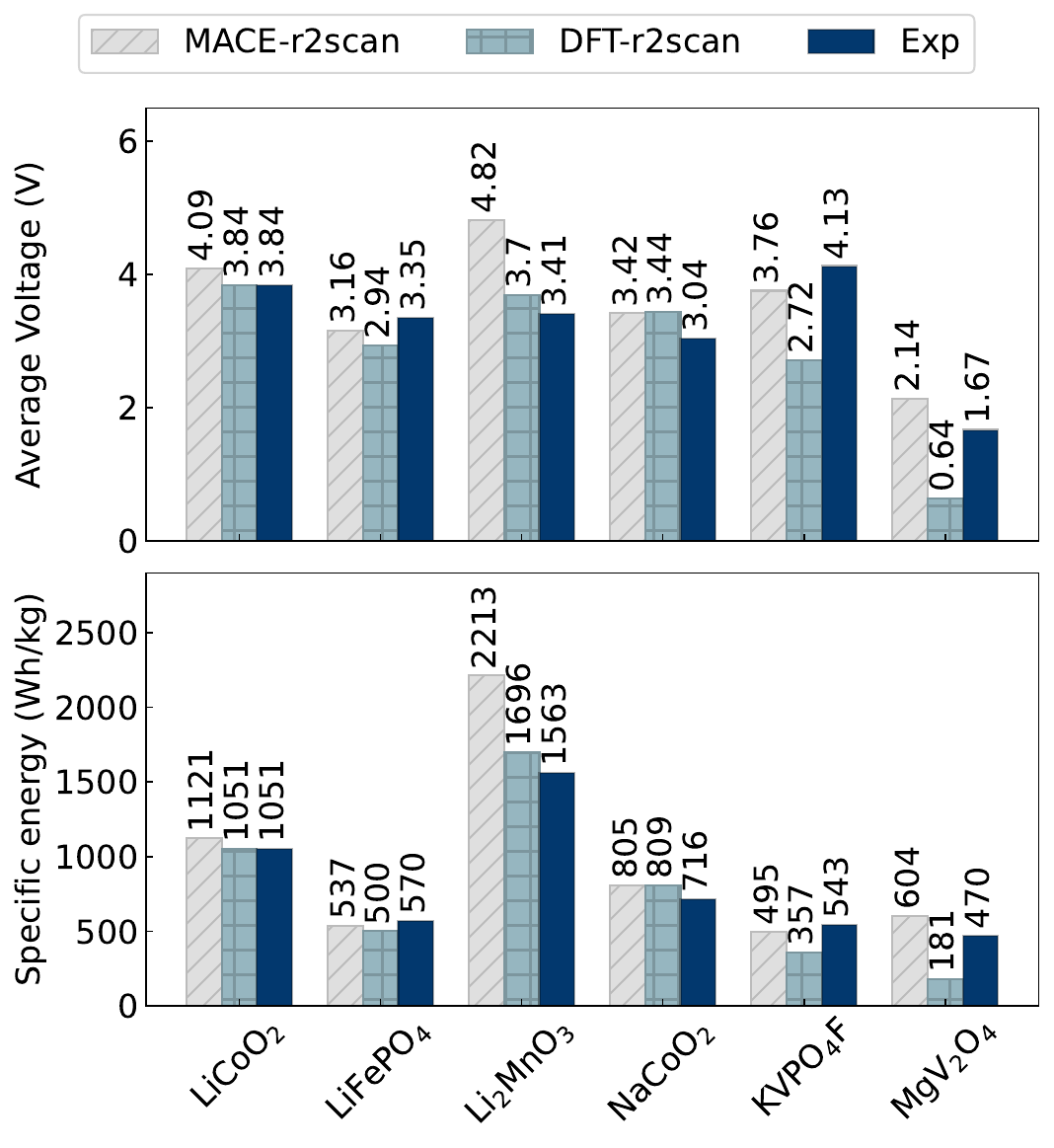}
    \caption{Comparison of the average voltages and specific energies obtained by MACE-r2scan, DFT-r2scan, and experiment. Experimental data were computed from the voltage profiles: \ce{LiCoO2}~\cite{HU202361}, \ce{LiFePO4}\cite{ROWDEN202197}, \ce{Li2MnO3}\cite{Rana2014Structural}, \ce{NaCoO2}\cite{reddy2015high}, \ce{KVPO4F}\cite{K-ion-cathode}, \ce{MgV2O4}\cite{Hu2020High}}
    \label{fig:comp_ave_V_E_density}
\end{figure}
Fig. \ref{fig:comp_ave_V_E_density} shows a comparison between the MACE-r2scan, DFT-r2scan, and experimental average voltages and specific energies for all the considered cathodes.
MACE-r2scan maintains consistent performance against the reference values, with the voltage values falling within $\pm 1.41$ V of the experimental reference.
The DFT-r2scan results are significantly worse than the MACE-r2scan in the \ce{MgV2O4} and \ce{KVPO4F} cathodes. 
This can be attributed to the omission of magnetic ordering in the DFT calculation.
Due to implementation limitations in Quantum Espresso at the time of running the calculations (v.7.4.1), the r$^2$SCAN functional calculations were performed with non-spin-polarized (paramagnetic) calculations.
\ce{MgV2O4} is known to have intricate spin-orbital coupling \cite{spin2023Lane}, that requires careful examination to determine the correct magnetic order of the ground state. 
On the other hand, \ce{KVPO4F} has been studied using DFT+U by Kim et al.~\cite{K-ion-cathode}.
The authors found that a high Hubbard U value of 5 eV was needed to get good agreement with the experimental results.
Using lower U values led to an underestimation of voltage values.

For the specific energy, the voltage profiles and theoretical specific capacities have been used to estimate the values, the results are again consistent, although \ce{Li2MnO3} is an outlier, this might be related to the co-existence of two phases of \ce{Li2MnO3} and \ce{LiMn2O4} at high voltages~\cite{inorganics2010132}.
In our method, phase stability is evaluated assuming framework-preserving intercalation/deintercalation processes (topotactic transition) and neglects possible reconstructive phase transitions.
Note also that there might be some differences between the theoretical possible and practical accessible  specific energy due to instability at higher rate of de-intercalation, phase transformation, decomposition at the electrode, among other reasons \cite{Meng2013Recent}.
That being said, the filtering criterion adopted for the specific energy is loose, we remove candidates only if the specific energy is below 250 Wh/Kg.
Importantly, it is worth noticing that MACE-r2scan can consistently discriminate the hierarchical order of the examined cathode materials in terms of the predicted specific energy.

\subsubsection{Phonon based dynamical stability}
Evaluating phonon dispersions of hypothetical materials is a widely used method in high-throughput calculations to filter candidates that are not stable (e.g., 
\cite{Malyi2019Energy, Wang2022High, Singh2018High}). 
Imaginary phonon modes signal dynamical instability, whereas positive real frequencies indicate that the system lies at a minimum of the potential energy surface (see Section \ref{ssec:phonon}).
A recent benchmark of foundational MLFF demonstrated a very good potential for predicting harmonic phonon properties \cite{loew2025universal}. 
The MACE-MP-0 \cite{Batatia2025foundation}, predicted dynamical stability with 95\% true positives in predicting stable structures and 73\% true unstable for predicting instability \cite{loew2025universal}.
However, the phonon dispersions of nearly 10,000 compounds used in this study \cite{loew2025universal, NIMS_MDR_2024} were computed on structures that are also present in the MPTrj dataset \cite{deng2023chgnet} which constituted the training data of the MACE-MP-0 \cite{Batatia2025foundation}.
In addition, the methodology of ref. \cite{loew2025universal} is such that they compute the phonon properties by optimizing the structures with the MLFF on top of the DFT relaxed geometries.
This might lead to overestimation of the accuracy of the MACE-MP-0 in detecting stability.

To this end, we test MACE-OMAT medium model on the same phonon dataset of Ref.~\cite{loew2025universal}. MACE-OMAT was trained on the OMat24 dataset~\cite{barrosoluque2024openmaterials2024omat24}, that contains more than 100 million DFT datapoint in equilibrium and out of equilibrium.
OMat24 is not directly dependent on or built from the MPTrj dataset. Therefore, while we cannot confirm that its 100 million data points are completely distinct from those in the phonon dataset, its independent construction stands as a better alternative. 
%
%
Moreover, the larger dataset size and the inclusion of out of equilibrium data provides more natural choice for phonon calculations \cite{Batatia2025foundation}.
Additionally, phonon calculations performed with MACE-OMAT do not include non-analytical corrections (NAC), which are responsible for splitting between longitudinal acoustic mode and transverse acoustic modes at the Brillouin-zone center. 
This approximation is acceptable for the present screening purposes, as dynamical stability is governed by the acoustic phonon branches, whereas NAC primarily affects high-frequency optical modes. Furthermore, NAC evaluation requires Born effective charges and dielectric tensors, which are not currently available within foundational MLFF frameworks.

\begin{figure*}[!ht]
    \centering
    \includegraphics[width=\linewidth]{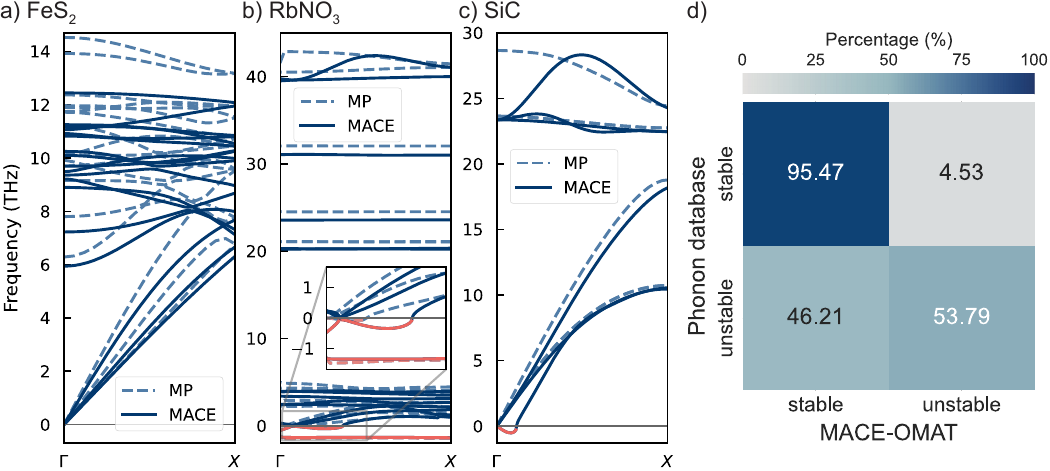}
    \caption{
            Phonon-based dynamical stability analysis.
            Phonon dispersions along the $\Gamma$-$X$ high-symmetry path are shown for (a) $\mathrm{FeS_2}$, (b) $\mathrm{RbNO_3}$, and (c) $\mathrm{SiC}$, comparing Materials Project reference calculations (dashed lines) with MACE-OMAT predictions (solid lines). Panel (d) presents the corresponding confusion matrix for dynamical stability classification, evaluated on a benchmark set of 7,360 stable and 1,585 unstable structures from the phonon database reported in Ref.~\cite{loew2025universal}.
        }
    \label{fig:confusion}
\end{figure*}

Fig.~\ref{fig:confusion} presents three representative phonon dispersions computed along the $\Gamma$-$X$ high-symmetry path for $\mathrm{FeS_2}$, $\mathrm{RbNO_3}$, and $\mathrm{SiC}$.
For $\mathrm{FeS_2}$ (Fig.~\ref{fig:confusion}.a) and $\mathrm{RbNO_3}$ (Fig.~\ref{fig:confusion}.b), MACE-OMAT reproduces the qualitative behavior of the acoustic phonon dispersions and correctly predicts dynamical stability and instability, respectively, in agreement with reference data.
Conversely, the $\mathrm{SiC}$ (Fig.~\ref{fig:confusion}.c) case highlights a limitation of the approach, where an imaginary acoustic mode leads to a false prediction of dynamical instability, despite otherwise good agreement for the remaining branches.
The performance of the MACE-OMAT against the reference \cite{loew2025universal, NIMS_MDR_2024} is shown in the confusion matrix Fig.~\ref{fig:confusion}.d.
Namely, MACE-OMAT achieved a 95\% true positive rate in predicting stable structures and a 54\% true negative rate for unstable ones.
This is consistent with the potential energy surface (PES) softening observed in many MLFF that leads to the softening of the phonon modes \cite{deng2025systematic, kaplan2025foundationalpotentialenergysurface} and overestimating the dynamical stability.
Nonetheless, this is acceptable for screening, as the stability of structures are estimated with accuracy of 88.13\%.

Remarkably, large supercells containing hundreds to thousands of atoms were used (details in the supplementary data), phonons were found with the finite displacement method, where each calculation was taking minutes using a single NVIDIA A100 GPU.
The DFT phonons calculations, whether using supercells or density functional perturbation theory method, is much more demanding and computationally expensive \cite{Bartel2022}.
The trade off between computational time and accuracy for using MACE-OMAT is exceptional for screening applications.

\subsection{Screening Energy-GNoME database}
\label{Screening Energy-GNoME database}
Our starting point is the identified cathodes in Energy-GNoME \cite{DEANGELIS2025100605, de_angelis_energy-gnome_2024}, with 90\% average confidence level based on the AI-experts committee (see Ref.~\cite{de_angelis_energy-gnome_2024} for details), and restricting the screening to cathodes relevant for Li-, Na-, K-, Mg- and Ca-ion batteries, which yields 615 cathode candidates (\textit{Filter 0} in Fig.~\ref{fig:workflow}).

\subsubsection{MACE-based and heuristics filtering}

At this stage, the screening protocol applies two filters grounded in explicit physical models, both accelerated by the use of foundational MLIPs. 
These filters aim to refine the initial Energy-GNoME candidate pool by incorporating thermodynamic and vibrational stability criteria that would otherwise require computationally prohibitive first-principles calculations.
The procedure starts with filtering out dynamically unstable candidates based on MACE-OMAT assessment.
We thus found 55 candidates (out of 615) that had imaginary phonons, those candidates are filtered out (\textit{Filter 1} in Fig.~\ref{fig:workflow}).

Next, based on the 
MACE-r2scan screening, we evaluate the average voltage values and specific energies by explicitly computing equilibrium intercalation voltage profiles at 0~K. 
MACE-r2scan is employed as a surrogate model for DFT, enabling the use of larger supercells and a finer convex-hull construction along the intercalation pathway at a fraction of the computational cost. 
Based on the voltage profiles, we applied a voltage window of 2.1-6~V, together with a minimum specific energy requirement of 250~Wh/kg (\textit{Filter 2} in Fig.~\ref{fig:workflow}).
This results in 269 remaining candidates (291 candidates were removed).

To further refine our candidate list, we employ two physically motivated heuristic filters based on space-group (SG) prevalence and elemental composition.
Experimental inorganic materials are not equally distributed among the 230 space groups, as shown in Fig. S3. 
In fact, two thirds of the known inorganic matter experimentally falls into only 24 space groups~\cite{urusov2009frequency}.
%
In the GNoME database, there is an overrepresentation of unlikely low symmetry SG, for example, 10.2\% of all GNoME structures have the $P1$ group (i.e. the group with the lowest symmetry, where only translational symmetry of the unit cell can be found) \cite{Cheetham2024Artificial}. 
However, in the Inorganic Crystal Structure Database (ICSD) searching for the $P1$ space group, result in only 813 entries out of the $\approx$ 241,000 inorganic crystals available (0.34\%) \cite{Zagorac2019ICED}.
In addition, non-centrosymmetric space groups are highly represented in the GNoME database, while in experimental synthesized inorganic matter, there is a preference for centrosymmetric crystals~\cite{Cheetham2024Artificial, HICKS2021110450, ECKERT2024112988}.
Therefore, as an indicator of structures with a higher likelihood of being synthesizable, we focus on candidates with space groups matching the 24 most frequently occurring space groups \cite{urusov2009frequency}.
Due to the overrepresentation of low-symmetry space groups inherited from GNoME and reflected in the Energy-GNoME distribution, only 93 of the 269 candidates pass this heuristic filter (\textit{Filter 3} in Fig.~\ref{fig:workflow}).

For the element composition of the candidates, we adopt the battery materials classification of Borah et al. \cite{BORAH2020OnBattery}, who classified elements based on abundance, cost, and toxicity. 
Hence, we categorize cathode candidates into three categories (\textit{Filter 4} in Fig.~\ref{fig:workflow}): 
\begin{description}
    \item[Class 1:] candidates composed entirely of feasible (abundant and inexpensive) elements (dark green flow in Fig.~\ref{fig:workflow});
    \item[Class 2:] candidates containing at least one marginally costly or marginally toxic element (green flow in Fig.~\ref{fig:workflow});
    \item[Class 3:] candidates containing at least one element with prohibitive cost or limited availability (orange flow in Fig.~\ref{fig:workflow}).
\end{description}
All candidates with acutely toxic/radioactive elements were removed (red flow in Fig.~\ref{fig:workflow}), leaving 89 candidates distributed among the three classes as follows: 9 in class~1, 35 in class~2, and 45 in class~3. 
The underlying rationale is that, for batteries as well as other technologies, it is desirable to achieve the highest possible performance using elements that are abundant and safe. 
The entire list of these candidates is presented in Tables. \ref{Table_1}, \ref{Table_2}, and \ref{Table_3}. 

Together, these physics-based and heuristic filters progressively reduce the candidate pool to a manageable set of materials that balance predicted electrochemical performance with physical plausibility and experimental feasibility.

\subsubsection{DFT refinement}

We perform DFT (PBEsol+U) screening on the remaining candidates only with feasible, marginally costly and minimally toxic elements (Tables. \ref{Table_1}, \ref{Table_2}).
We applied stricter voltage criteria, selecting only candidates within the 3-5 V range.
The lower bound ensures competitive performance, while the upper bound is imposed to avoid electrolyte decomposition~\cite{BORAH2020100046}.

\begin{figure*}[!ht]
    \centering
    \includegraphics[width=0.45\linewidth]{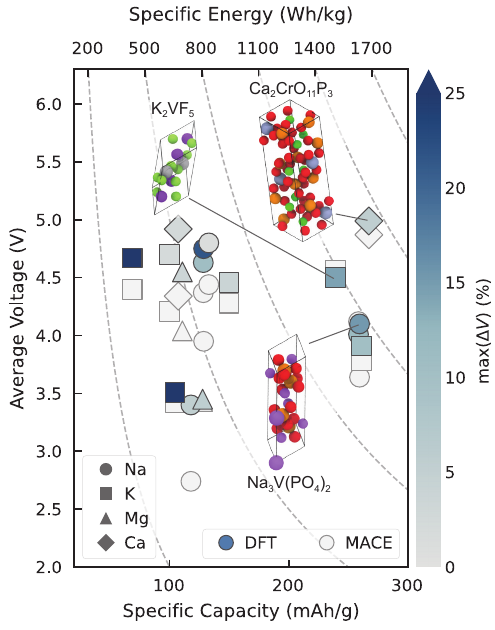}
    \caption{
            Post-lithium cathode candidates after DFT refinement.
            The scatter plot shows theoretical gravimetric capacity versus average voltage relative to the pure-element oxidation potential ($\mathrm{X/X^{n+}}$, with $\mathrm{X}$ denoting the working ion). Gray dashed hyperbolas correspond to constant gravimetric energy.
            For each cathode, both MACE and DFT predictions are reported.
            The latter are colored according to the maximum volume expansion, $\max(\Delta V)$.
            }
    \label{fig:FinalCandidates}
\end{figure*}

Finally, we add one physically motivated filter, because experimentally, most known inorganic materials contain two to five distinct elements.
In the ICSD \cite{Zagorac2019ICED}, 91\% of the entries contain between two and five elements and 77\% of the entries contain only two to four elements.
The higher the number of elements in the crystal, the more challenging it is to synthesize \cite{sun2025critical}.
Hence, in the last step we keep only the candidates with less than five elements.
The final shortlisted candidates are shown in Fig.~\ref{fig:FinalCandidates} and their detailed values are listed in Table. \ref{tab:shortlisted_candidates}.
Many of the final cathode candidates contain polyanionic units, including Phosphate (\ce{PO4}) and pyrophosphate (\ce{P2O7}).
These frameworks are known to provide structural stability and can host a variety of transition-metal redox couples, which is beneficial for reversible electrochemical performance.
A summary of the number of candidates through the screening process is shown in Table.~\ref{tab:summary_cathode_candidates}

\setcounter{table}{3}
\begin{table}[htpb]
\caption{Shortlisted candidates using the protocol in Fig. \ref{fig:workflow}.}\label{tab:shortlisted_candidates}
\centering
{\renewcommand{\arraystretch}{1.15}
    \begin{tabular}{cll|cc|cc|c}
    \hline 
    \multirow{2}{*}{\textbf{GNoME ID}}& \multirow{2}{*}{\textbf{Ion}} & \multirow{2}{*}{\textbf{Formula}} & \multicolumn{2}{c|}{$\mathbf{\overline{V}}$ (V)} & \multicolumn{2}{c|}{$\mathbf{e}$ (Wh/kg)} & $\mathbf{\max(\Delta V)}$ (\%) \\
    \cline{4-8}
    &  &  & \textbf{DFT}  & \textbf{MACE} & \textbf{DFT} & \textbf{MACE} &  \textbf{MACE} \\
    \hline 
    0a6ddde9d0 & Na & \ce{Na2Fe(PO3)4} & 4.63 & 4.45 &   594.15 & 570.94 & 1.92 \\
    5c2e6d9372 & Na & \ce{Na2Mn(PO3)4} & 4.75 & 4.24 & 610.88 &  546.20 & 7.96 \\
    35ca0bcad2 & Na & \ce{Na4Fe5(P3O11)2} & 3.40 & 2.77 & 400.99 & 326.70 & 0.72 \\
    400fab0a1d & Na & \ce{Na3Cr3(PO4)4} & 4.80 & 4.57 &  638.09 & 608.22 & 13.41 \\
    405e95bfd6 & Na & \ce{Na3Cr(PO4)2} & 4.01 & 4.46 & 1037.04 & 1154.11 & 16.07 \\
    949e559e66 & Na & \ce{Na3V(PO4)2} & 4.10  & 3.97 & 1063.93 & 1032.04 & 12.37 \\
    \hline
    45fd5fdb00 & Mg & \ce{MgFe2(P2O7)2} & 4.55 & 3.88 & 504.04 & 429.62 & 19.15 \\
    b08b4171d3 & Mg & \ce{Mg(FeCl4)2} & 3.45 & 3.23 & 440.74 & 413.02 & 36.45 \\
    \hline
    5a35b9a7d8 & K & \ce{K2Mn5O10} & 3.51 & 3.29 & 366.85 & 344.26 & 11.70 \\
    ab6e151f05 & K & \ce{K2VF5} & 4.50 & 4.39 & 1076.22 & 1049.20 & 9.23 \\
    bb0f344d09 & K & \ce{KMn3CoO8} & 4.67 & 4.39 & 320.24 &  301.13 & 7.91 \\
    dd00bcc6e3 & K & \ce{K2V(SiO3)3} & 4.46 & 4.28 & 668.94 & 641.77 & 6.37 \\
    75ae834ca3 & K & \ce{K2V(Si2O5)3} & 4.70 & 4.21 & 468.60 & 419.67 & 5.77 \\
    016d5b7e2b & K & \ce{K4F14CaV3} & 3.91 & 3.72 & 681.28 & 648.47 & 36.73 \\
    \hline
    1de51c5a90 & Ca & \ce{CaFe2(P2O7)2} & 4.92 & 4.42 & 527.82 & 474.48 & 12.11 \\
    6e82276b80 & Ca & \ce{Ca2CrO11P3} & 4.99 & 5.12 & 1333.85 & 1368.73 & 14.64 \\
    \hline
    \end{tabular}
}
\end{table}
\begin{table}[ht!]
\centering
\caption{The number of candidates at each stage of the screening process.}
\label{tab:summary_cathode_candidates}
\begin{tabular}{l|ccccc|c}
\hline
& \multicolumn{5}{c|}{Number of candidates} & \\
& Li-ion & Na-ion & K-ion & Mg-ion & Ca-ion & total \\
\hline
Starting dataset (filter 0) & 15 & 104 & 231 & 235 & 30 & 615 \\
MACE screening (filters 1 \& 2) & 9 & 39 & 86 & 113 & 22 & 269 \\
Physical heuristics (filters 3 \& 4) & 2 & 18 & 27 & 31 & 11 & 89 \\
Refined by DFT (Tables. \ref{Table_1} \& \ref{Table_2}) & 0 & 9 & 19 & 8 & 8 & 44 \\
Final shortlisted candidates & 0 & 6 & 2 & 6 & 2 & 16 \\
\hline
\end{tabular}
\end{table}

\clearpage
\subsection{Methods comparison}
\begin{figure*}[!ht]
    \centering
    \includegraphics[width=\linewidth]{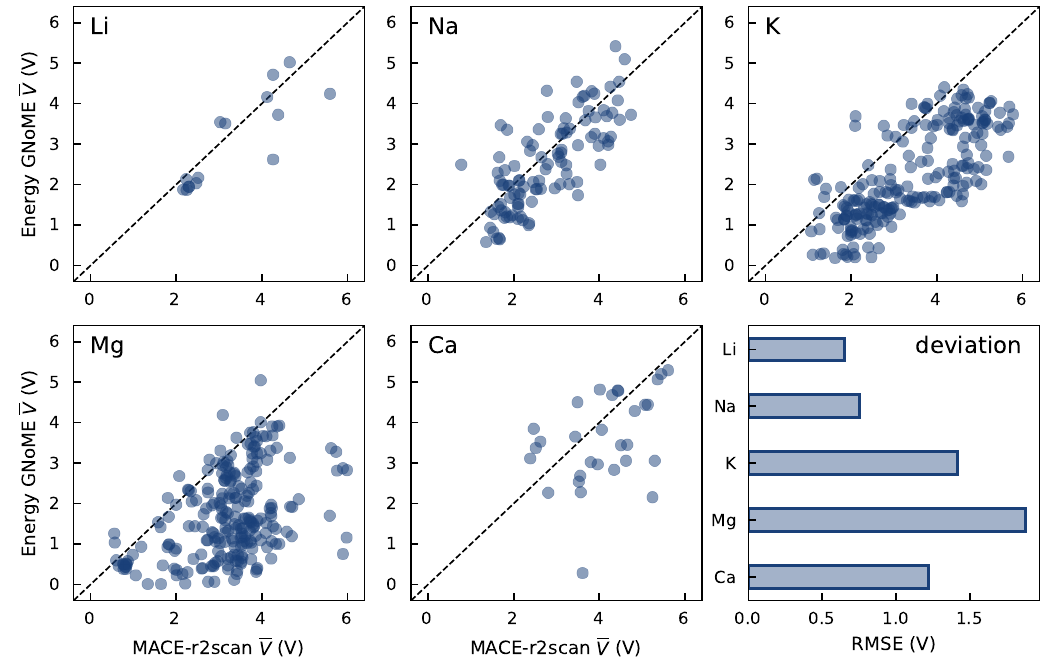}
    \caption{
            Comparison of the average voltage $\overline{V}$ predicted by MACE-r2scan against predictions from Energy-GNoME
    }
    \label{fig:Energy-GNoME_vs_MACE}
\end{figure*}

\begin{figure*}[!ht]
    \centering
    \includegraphics[width=\linewidth]{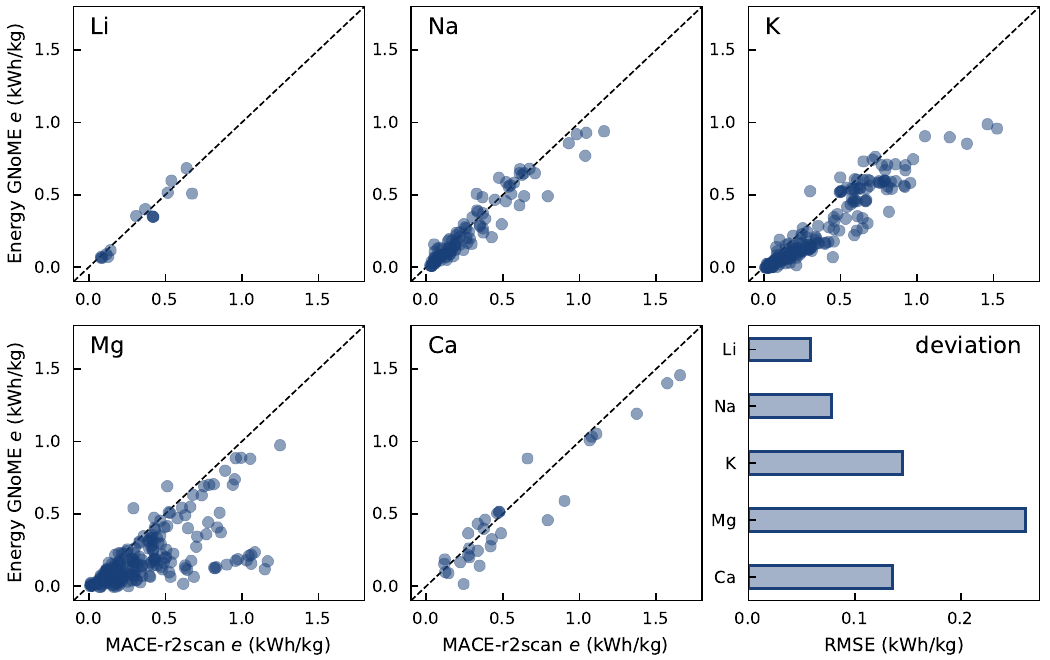}
    \caption{
            Comparison of the specific energy $e$ predicted by MACE-r2scan against predictions from Energy-GNoME.
    }
    \label{fig:Energy-GNoME_vs_MACE_energy}
\end{figure*}

\begin{figure}[!ht]
    \centering
    \includegraphics[width=0.45\linewidth]{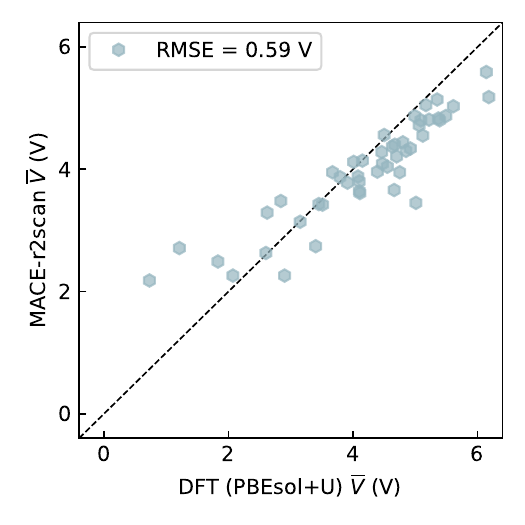}
    \caption{
        Comparison of average voltages $\overline{V}$ calculated with DFT (PBEsol+U) and MACE-r2scan, for the 44 cathode candidates from Tables \ref{Table_1} and \ref{Table_2}, together with 4 additional candidates. While the methods employ different levels of theory, MACE-r2scan shows good agreement with the DFT results.
    }
    \label{fig:DFT_vs_MACE}
\end{figure}

We present a comparison between the MACE prediction and the corresponding readily available values in the Energy-GNoME database \cite{de_angelis_energy-gnome_2024} and the DFT+U computed average voltage.
Fig. \ref{fig:Energy-GNoME_vs_MACE} shows a fairly good correlation between the MACE-r2scan and the Energy-GNoME reported values for the average voltage.
The root mean square error (RMSE) is highest for Mg. This can be attributed to the fact that Energy-GNoME predictions rely on a pure regression model,  and although the dataset for Mg contained about 423 data points, it was, however, an external contribution to the Materials Project (MP), so it may not be as reliable as standard MP data.
Next, K-ion exhibits the second highest RMSE value.
This correlates with the size of the training datasets used for the K-ion, which included only 92 data points.
By comparison, the datasets for Na-, Ca-, and Li-ion systems include 309, 416, and 2272 data points, respectively.
Similar trends can be observed for the specific energies in Fig. \ref{fig:Energy-GNoME_vs_MACE_energy}.
On the other hand, although the level of theory behind the DFT (PBEsol+U) and the MACE-r2scan model is different, it is still meaningful to check the performance of MACE against DFT. Fig. \ref{fig:DFT_vs_MACE}, shows the parity plot of the average voltage, with good agreement between the two methods and RMSE of 0.59~V. See Table S2 in the Supplementary Information for detailed values.

\clearpage
\section{Method}
\subsection{Convex hull and voltage profile}
\label{method:Convex_hull_and_voltage_profile}
    The thermodynamics of ion intercalation and deintercalation in a generic cathode active material can be described by the following electrochemical reaction between a working ion $\mathrm{A}$ (e.g., Li, Na, Mg) and a host framework $\mathrm{B}$:
    \begin{equation}
    \label{eq:intercalation_reaction}
    x\,\mathrm{A}^{n+} + x\,n\,e^- + \mathrm{B}
    \;\rightleftharpoons\;
    \mathrm{A}_{x}\mathrm{B},
    \end{equation}
    where $n$ denotes the number of electrons transferred per intercalated ion (e.g. $n_{\mathrm{Li}}=1$, $n_{\mathrm{Mg}}=2$).
    The formation energy $\Delta E_f$ for the generic cathode active material $\mathrm{A_{x}B}$ can be expressed as~\cite{Urban2016}:
    \begin{equation}
    \label{eq:formation_energy}
    \Delta E_{f}\left(\mathrm{A}_{x}\mathrm{B}\right)
    =
    E\left(\mathrm{A}_{x}\mathrm{B}\right)
    -
    x E\left(\mathrm{A}\mathrm{B}\right)
    -
    (1-x) E\left(\mathrm{B}\right),
    \end{equation}
    where $0 \le x \le 1$ is the fractional occupation of intercalation sites,  $E(\mathrm{A}_{x}\mathrm{B})$ is the total internal energy per formula unit of the host framework at composition $x$, $E(\mathrm{B})$ is the reference energy of the pristine (fully de-intercalated) host framework, and $E(\mathrm{A}\mathrm{B})$ represents the energy of the fully intercalated reference phase.

The potential energies 
$E$ of each phase are either MACE predicted energy or DFT computed energy.
This quantity describes the relative stability of the intermediate phase 
$\mathrm{A}_{x}\mathrm{B}$, with respect to the fully intercalated 
$\mathrm{A}\mathrm{B}$ and de-intercalated 
$\mathrm{B}$ phases, where thermodynamically stable phases lie on the convex hull of the composition-energy ($x$ vs. 
$\Delta E_f$) diagram (Fig.~\ref{fig:convex_hull_ave_V_LiCoO2}).

    The voltage profile $V(x_1,x_2)$ is computed as a piecewise constant approximation between two adjacent stable compositions $x_1 > x_2$ along the intercalation pathway as~\cite{Urban2016}:
    \begin{equation}
    \label{eq:voltage_steps}
    V(x_1,x_2) 
    = 
    -\dfrac{\Delta G(x_1)-\Delta G(x_2)}{nF(x_1-x_2)}
    \approx
    -\dfrac{
    E\left(\mathrm{A}_{x_1}\mathrm{B}\right)
    -
    E\left(\mathrm{A}_{x_2}\mathrm{B}\right)
    -
    (x_1-x_2)\,E(\mathrm{A})
    }{
    n F (x_1-x_2) 
    },
    \end{equation}
    where $\mathrm{A}_{x_1}\mathrm{B}$ and $\mathrm{A}_{n x_2}\mathrm{B}$ are thermodynamically stable phases, $n$ is the number of electrons transferred per intercalated ion, $F$ is the Faraday constant,and $E(\mathrm{A})$ is the reference energy of the working-ion reservoir.
At low temperatures, entropic contributions are negligible, and the energies $E$ can be equated to the internal energy.
Once the voltage profile is constructed, the average voltage value can be determined by averaging the voltage profile and the specific energy $e$ can be obtained from the average voltage and the theoretical maximum capacity.
Alternatively, to find the average voltage, $\overline{V}$, for a given cathode, it is enough to compute the energy of the fully intercalated phase, fully de-intercalated phase, and the chemical potential of the working ion \cite{Urban2016}:
\begin{equation}
\label{eq:average_voltage}
\overline{V}  = V(1, 0)\approx
\dfrac{
 -E\left(\mathrm{A}\mathrm{B}\right)
+E\left(\mathrm{B}\right)
+ E(\mathrm{A})
}{
nF
}
\end{equation}
The maximum volumetric expansion is determined as the relative change between the maximum and minimum volumes:
\begin{equation}
\max (\Delta V) = \frac{V_{\max} - V_{\min}}{V_{\min}}
\end{equation}


\subsection{Phonon-based dynamical stability}\label{ssec:phonon}
The dynamical stability of a crystal structure can be assessed through its vibrational properties.
To this end, the potential energy $V$ is expanded as a Taylor series around the equilibrium atomic positions~\cite{AshcroftMermin}:
\begin{equation}
V\left(\left\{\mathbf{R}^0+\Delta \mathbf{R}\right\}\right) \approx 
V\left(\left\{\mathbf{R}^0\right\}\right)+
\left.\sum_J \frac{\partial V}{\partial \mathbf{R}_I}\right|_{\mathbf{R}^0} \Delta \mathbf{R}_I
+\left.\frac{1}{2} \sum_{I, J} \frac{\partial^2 V}{\partial \mathbf{R}_I \partial \mathbf{R}_J}\right|_{\mathbf{R}^0} \Delta \mathbf{R}_I \Delta \mathbf{R} 
+\mathcal{O}\left(\mathbf{R}^3\right)
\end{equation}
where $\Delta\mathbf{R}$ denotes a  small displacement from the equilibrium positions $\mathbf{R}^0$, and the summation runs over atoms $I$ and $J$.
At equilibrium, the forces $\mathbf{F}_I$ acting on each atom vanish, i.e.
\begin{equation}
\mathbf{F}_I \equiv-\frac{\partial V(\mathbf{R})}{\partial \mathbf{R}_I}=0,
\end{equation}
The second-order term in the expansion defines the Hessian of the potential energy, also known as the interatomic-force constants matrix:
\begin{equation}
\label{eq:force_constant}
\Phi_{I J}=\left.\frac{\partial^2 V}{\partial \mathbf{R}_I \partial \mathbf{R}_J}\right|_{\mathbf{R}^0}=-\left.\frac{\partial}{\partial \mathbf{R}_I} \mathbf{F}_J\right|_{\mathbf{R}^0}
\end{equation}
within the harmonic approximation, the equation of motion for lattice vibration leads to the following eigenvalue problem:
\begin{equation}
\mathbf{D}(\mathbf{q})\ \boldsymbol{\epsilon}_\nu(\mathbf{q})
=\omega^2_\nu(q) \boldsymbol{\epsilon}_\nu(\mathbf{q})
\end{equation}
where $\mathbf{q}$ are the phonon wave vectors, $\omega_\nu(q)$ are the frequencies of the phonon modes, and $\mathbf{D}$ is the dynamical matrix, which is constructed by a mass-weighted Fourier transform of the interatomic-forces
\begin{equation}
D_{I\alpha J \beta}(\mathbf{q})=
\frac{1}{\sqrt{m_I m_J}} 
\sum_\mathbf{R} \Phi_{I \alpha, J \beta} e^{-i \mathbf{q}\cdot\mathbf{R}},
\end{equation}
%
%
A structure is dynamically stable if all phonon frequencies are real and positive, indicating that the system resides at a minimum of the potential energy surface. In contrast, the presence of imaginary phonon modes signals dynamical instability \cite{AshcroftMermin, pallikara2022physical}.


\subsection{Computational framework}
%
As many cathode systems depends on transition metal (TM) elements as electrochemical active elements, whose oxidation states change during charge–discharge cycling, the use of simple semi-local functionals is not an optimal choice \cite{Zhou2004First, Accurate2022Timrov}.
Semi-local DFT functionals suffer from electron self-interaction error (SIE), which becomes particularly dominant in systems with strongly correlated electrons, such as TMs with open $d$-shells~\cite{Kulik2008self}.
%
The Hubbard correction U, has been shown to mitigate the SIE and improve intercalation-voltage prediction \cite{Zhou2004First, Accurate2022Timrov}. 
In addition, meta-GGA functional exhibit reduction of SIE compared to semi-local functionals, and generally achieves better accuracy for TMs systems at a comparable computational cost~\cite{zhang2025advanceschallengesscanr2scan}.
To this end, we used spin-polarized PBEsol+U for the final evaluation of the average voltage of the most promising cathode candidates, while the non-spin-polarized r$^2$SCAN meta-GGA functional was used in the context of our initial rationalization of the validity of our approach against the experimental intercalation voltage profile.

%

%
Moreover, for screening, we employed two foundational machine learning force fields models based on the MACE architecture \cite{Batatia2022mace}: MACE-MATPES-r2SCAN-0 model (MACE-r2scan for short), and MACE-OMAT-0 (MACE-OMAT)~\cite{mace-foundations}.
The MACE-r2scan was trained on DFT data at the r$^2$SCAN functional level using the MATPES-r2scan dataset \cite{kaplan2025foundationalpotentialenergysurface} which includes 387,897 structures both at equilibrium and non-equilibrium. 
While the MACE-OMAT was trained using the OMat24 dataset comprised of over 110 million DFT which is sampled from a diverse non-equilibrium data points~\cite{barrosoluque2024openmaterials2024omat24}.
It relied on the PBE functional with Hubbard U correction for Co, Cr, Fe, Mn, Mo, Ni, V, or W transition metal elements in oxides and fluorides \cite{mace-foundations, barrosoluque2024openmaterials2024omat24}.
The different methods and the way we used them are summarized in Table.~\ref{tab:methods}.

\begin{table}[!ht]
    \centering
    \caption{Summary of the machine learning interatomic potentials (MACE) and density functional theory (DFT) methods used in this work.}\label{tab:methods}
    {\renewcommand{\arraystretch}{1.15}
    \begin{tabular}{c c c m{2.7cm} m{4cm}}
        \hline
        \textbf{Method} & \textbf{Model/Functional} & \textbf{Level of Theory} & \textbf{Usage} & \textbf{Notes}\\
        \hline
        \multirow{2}{*}[-3.2ex]{\centering MACE} & MATPES-r2SCAN-0 & meta-GGA  &Voltage profiles and average values (eqs. \ref{eq:formation_energy}, \ref{eq:voltage_steps}) & Trained on MATPES-r$^2$SCAN dataset \cite{kaplan2025foundationalpotentialenergysurface}  \\
        \cline{2-5}
                              & OMAT-0           & GGA+U & Phonon-based dynamical stability (eq. \ref{eq:force_constant}) & Trained on OMat24 dataset \cite{barrosoluque2024openmaterials2024omat24}, with PBE functional and Hubbard U applied to Co, Cr, Fe, Mn, Mo, Ni, V and W, in oxides and fluorides \cite{barrosoluque2024openmaterials2024omat24}.\\
        \hline
        \multirow{2}{*}[-3.5ex]{\centering DFT}  &  r$^2$SCAN  & meta-GGA & Voltage profiles and average values (eqs. \ref{eq:formation_energy}, \ref{eq:voltage_steps}) & Non-spin polarized calculations \\
        \cline{2-5}
                              & PBEsol+U   & GGA+U & Average voltages (eq. \ref{eq:average_voltage}) & Spin-polarized calculations and fixed Hubbard correction values $U_{Ni}=5.8$, $U_{Mn} = 4.7$, $U_{Fe} = 4.5$, $U_{V} = 3.9$, $U_{Cr} = 3.0$, $U_{Co} = 5.2$, and $U_{Ti} = 4.7$ eV \cite{Moore2024High}. \\
        \hline
    \end{tabular}
    }
\end{table}

\subsection{Technical setup}
%
Density functional theory (DFT) calculations were performed using Quantum ESPRESSO~\cite{giannozzi2009quantum, Giannozzi_2017, Giannozzi2020exascale} interfaced with the Atomic Simulation Environment (ASE) \cite{HjorthLarsen_2017}. 
Structural relaxations were performed via variable-cell relaxation (vc-relax).
For the final cathode candidates voltage refinement, a spin polarized calculation with the PBEsol exchange and correlation functional \cite{Perdew2008Restoring, Prandini2018} and Hubbard correction \cite{Dudarev1998electron, Cococcioni2005LR} was used.
We employed the SSSP PBEsol library (version 1.3)~\cite{prandini2018precision} with the default suggested cutoff values.
The starting inputs were generated with the Quantum ESPRESSO input generator utility hosted on Materials Cloud~\cite{mounet2018materialscloud}. 
A balanced protocol was chosen with k-points distance of 0.15 \AA$^{-1}$ and a smearing value of 0.02 Ry. An intermediate convergence threshold of $10^{-5}$ Ry/atoms for the total energy and $10^{-4}$ Ry/bohr for the forces and $2\times10^{-10}$ Ry/atom for the self-consistency cycle.
Hubbard corrections were included in the calculations within the rotationally invariant formalism of Dudarev et al.~\cite{Dudarev1998electron} as implemented in Quantum ESPRESSO~\cite{Giannozzi_2017, Giannozzi2020exascale}. 
Fixed Hubbard values, corresponding to the average values obtained from a high-throughput evaluation of 1000 TM systems were employed~\cite{Moore2024High}.
Namely, $U_{Ni} = 5.8$, $U_{Mn} = 4.7$, $U_{Fe} = 4.5$, $U_{V} = 3.9$, $U_{Cr} = 3.0$, $U_{Co} = 5.2$, and $U_{Ti} = 4.7$ eV.

%
In the method validation section, the r$^2$SCAN meta-GGA functional \cite{r2scan} was used for the exchange-correlation functional. The r$^2$SCAN was used as implemented in the Libxc library \cite{LEHTOLA20181}, in conjunction with Optimized Norm-Conserving Vanderbilt (ONCV) pseudopotentials \cite{PhysRevB.88.085117, SCHLIPF201536}. 
The following convergence thresholds were chosen: Forces: $10^{-3}$ Ry/Bohr, energy: $10^{-4}$ Ry and scf self-consistency: 10$^{-8}$ Ry.
The k-point meshes and plane-wave cutoff energies, in this case were selected to ensure total energy convergence within $10^{-2}$--$10^{-3}$~Ry.
The optimized parameters used for each system are summarized in Table. S3.

MLFF screening performed using MACE (version 0.3.13) \cite{Batatia2022mace, Batatia2022Design} in combination with ASE~\cite{HjorthLarsen_2017}. 
The structural optimization was performed using the FIRE2 and BFGS optimizers with a force convergence criterion of $10^{-4}$ eV\AA \ in both the ion intercalation steps and phonon evaluation.

To evaluate the dynamical stability of the structures, the supercell finite difference approach was used with the ASE phonons module.
A tolerance of $4\times10^{-3}$ eV was considered to filter out numerical artifacts in the phonon spectrum.
To evaluate the performance of MACE prediction of dynamical stability, we used the following metrics.
The true positive rate (also known as recall or sensitivity):
\begin{equation}
\mathrm{True\ positive\ rate} = \frac{\mathrm{TP}}{\mathrm{TP} + \mathrm{FN}} \, .
\end{equation}
The true negative rate (specificity):
\begin{equation}
\mathrm{True\ negative\ rate} = \frac{\mathrm{TN}}{\mathrm{TN} + \mathrm{FP}} \, .
\end{equation}
The accuracy:
\begin{equation}
\mathrm{Accuracy} = \frac{\mathrm{TP} + \mathrm{TN}}{\mathrm{TP} + \mathrm{TN} + \mathrm{FP} + \mathrm{FN}} \, .
\end{equation}
where $\mathrm{TP}$/$\mathrm{TN}$ denotes true positives/negatives, i.e., structures predicted to be stable/unstable and indeed stable/unstable.
On the other hand, $\mathrm{FP}$/$\mathrm{FN}$ are false positives/negatives, i.e., unstable/stable structures incorrectly predicted to be stable/unstable.

To determine the space groups of the structures, we used the \texttt{check\_symmetry} method from the ASE \texttt{spacegroup} subpackage, which interfaces with the \texttt{Spglib} library \cite{Togo31122024}.
A symmetry tolerance, i.e., the Cartesian distance tolerance to determine the crystal symmetry, of $10^{-4}$ was 
employed.
Only the most 24 frequent space groups are retained. These are: $P\overline{1}$, $P2_1/m$, $C2/m$, $P2_1/c$, $C2/c$, $Pna2_1$, $Pmmm$, $Pbca$, $Pnma$, $Cmcm$, $P4/nmmm$, $I4/mmm$, $R\overline{3}$, $P\overline{3}m1$, $R\overline{3}m$, $R\overline{3}c$, $P6_3/m$, $P6/mmm$, $P6_3/mmc$, $F\overline{4}3m$, $Pm\overline{3}m$, $Fm\overline{3}m$, and $Fd\overline{3}m$ \cite{urusov2009frequency}.

\section{Conclusions}
In this work, we started from the predictions on novel cathode materials that are readily available in the Energy GNoME database \cite{de_angelis_energy-gnome_2024} and applied a 
hybrid screening procedure based on 
multi-level theoretical framework consisting of both foundational MACE and DFT simulations to enable scalable filtering of very large materials databases.
Furthermore, on the basis of expected stability, typical space groups of experimentally known materials, toxicity, cost and expected synthesis ease, we have shortlisted a dozen of novel material candidates spanning Na-ion, Mg-ion, K-ion and Ca-ion chemistry.
It is worth stressing that this study provides a first clear evidence that the readily available figures of merit of the Energy-GNoME database, although only based on ML prediction, are comparable to MACE and DFT quality results.
Moreover, the present study represents the first but key step towards an {\it in silico} discovery of new cathode materials to be tested experimentally.
This is particularly timely, given the recent advances in generative computational materials discovery, which can rapidly produce vast numbers of candidate materials but still require physically grounded and interpretable screening strategies~\cite{DeBreuck2025GenerativeReview, zeni2025generative, Mikkel2026Continued}.
Among other things, a key aspect to predict would be the transport properties of the active material in the new material candidates.
Such a screening will be the subject of future studies, where molecular dynamics simulations can be adopted as demonstrated in \cite{alghamdi2025comparing,de2024enhancing,de2025exploring}.

\section{Author Contributions}

CRediT: Conceptualization: NA, PDA, PA, EC; Data curation: NA, PDA; Formal Analysis: NA, PDA; Funding acquisition: PA, EC; Investigation: NA; Methodology: NA, PDA, PA, EC; Project administration: PA, EC; Resources: PA, EC; Software: NA, PDA; Supervision: PDA, PA, EC; Visualization: NA, PDA; Writing - original draft: NA; Writing - review \& editing: NA, PDA, PA, EC
\section{Acknowledgment}

We acknowledge INRiM's  Better measurements for energy storage (\textit{Sviluppo di nuova metrologia e nuovi sensori per l'accumulo di energia elettrica per la transizione energetica e l'economia circolare}) project funded by Ministry of Research and University – \textit{MUR Progettualità di carattere continuativo}.
We also thank ISCRA (\verb|IsB29_NEXT-LIB|) for awarding access to the LEONARDO supercomputer, owned by the EuroHPC Joint Undertaking, hosted by CINECA (Italy).
\section{Conflict of interest}
The authors declare no conflicts of interest.
\section{Data availability}
The data will be made publicly available at \url{https://github.com/AlghamdiNada/screening_cathodes_Energy-GNoME}.

\setcounter{table}{0}

\begin{sidewaystable}[!htbp]
\centering
\caption{Candidates from the MACE screening with feasible elements following Ref. \cite{BORAH2020OnBattery} classification.}\label{Table_1}
\begin{tabular}{llll|cccc|cccc|l}
\hline
& & & & \multicolumn{4}{|c|}{MACE-r2scan} & \multicolumn{4}{c|} {Energy-GNoME prediction} &  \\
 \cline{5-12}
GNOME id & ion & cathode & space group & V$_{\mathrm{avg}}$  & E$_{\mathrm{grav}}$ & E$_{\mathrm{vol}}$ & $\Delta$ V & V$_{\mathrm{avg}}$ & E$_{\mathrm{grav}}$ & E$_{\mathrm{vol}}$ & $\Delta$ V  & Elements \\
 &  & & &(V) & (Wh/kg) & (Wh/L) & \% & (V) & (Wh/kg) & (Wh/L) & \% & feasibility \\
\hline
0a6ddde9d0 & Na & \ce{Na2Fe(PO3)4} & 15 (C2/c) & 4.45 & 570.94 & 1500.45 & 1.92 & 4.54 & 582.74 & 1531.49 & 2.16 &  \\
5c2e6d9372 & Na & \ce{Na2Mn(PO3)4} & 15 (C2/c) & 4.25 & 546.20 & 1410.83 & 7.96 & 4.42 & 568.37 & 1468.09 & 2.84 &  \\
35ca0bcad2 & Na & \ce{Na4Fe5(P3O11)2} & 14 (P2$_{1}$/c) & 2.77 & 326.70 & 1092.82 & 0.72 & 4.31 & 508.79 & 1701.88 & 4.98 &  \\
5a35b9a7d8 & K & \ce{K2Mn5O10} & 2 (P$\overline{1}$) & 3.29 & 344.26 & 1179.15 & 11.70 & 1.97 & 205.80 & 704.92 & 18.69 &  \\
45fd5fdb00 & Mg & \ce{MgFe2(P2O7)2} & 62 (Pnma) & 3.88 & 429.62 & 1192.90 & 19.15 & 3.57 & 394.97 & 1096.68 & 2.13 &  \\
60492c51e2 & Mg & \ce{MgFe3(NO3)8} & 2 (P$\overline{1}$) & 3.45 & 268.56 & 687.27 & 1.16 & 1.48 & 115.27 & 294.98 & 3.95 &  \\
b08b4171d3 & Mg & \ce{Mg(FeCl4)2} & 15 (C2/c) & 3.23 & 413.02 & 785.87 & 36.45 & 2.25 & 287.82 & 547.64 & 4.97 &  \\
2ab93c60bd & Ca & \ce{Ca3Fe(PO3)8} & 2 (P$\overline{1}$) & 5.34 & 1063.15 & 2789.99 & 5.42 & 5.07 & 1010.10 & 2650.77 & 8.63 &  \\
1de51c5a90 & Ca & \ce{CaFe2(P2O7)2} & 2 (P$\overline{1}$) & 4.42 & 474.48 & 1500.70 & 12.11 & 4.80 & 515.44 & 1630.25 & 2.16 &  \\
\hline
\end{tabular}
\end{sidewaystable}

\begin{sidewaystable}
\centering
\small
\caption{Candidates from the MACE screening with at least one elemnt of marginal cost/toxicity following Ref. \cite{BORAH2020OnBattery} classification, where MC: marginal cost and MT: marginal toxicity.}\label{Table_2}
\begin{tabular}{llll|cccc|cccc|l}
\hline
& & & & \multicolumn{4}{|c|}{MACE-r2scan} & \multicolumn{4}{c|} {Energy-GNoME prediction} &  \\
 \cline{5-12}
GNOME id & ion & cathode & space group & V$_{\mathrm{avg}}$  & E$_{\mathrm{grav}}$ & E$_{\mathrm{vol}}$ & $\Delta$ V & V$_{\mathrm{avg}}$ & E$_{\mathrm{grav}}$ & E$_{\mathrm{vol}}$ & $\Delta$ V  & Elements \\
 &  & & &(V) & (Wh/kg) & (Wh/L) & \% & (V) & (Wh/kg) & (Wh/L) & \% & feasibility \\
\hline
400fab0a1d & Na & \ce{Na3Cr3(PO4)4} & 15 (C2/c) & 4.58 & 608.22 & 1895.57 & 13.41 & 5.09 & 676.61 & 2108.71 & 3.39 & Cr:MT,  \\
405e95bfd6 & Na & \ce{Na3Cr(PO4)2} & 15 (C2/c) & 4.46 & 1154.11 & 3652.60 & NaN & 3.61 & 933.40 & 2954.08 & 14.76 & Cr:MT,  \\
949e559e66 & Na & \ce{Na3V(PO4)2} & 15 (C2/c) & 3.98 & 1032.04 & 3199.70 & 12.38 & 2.96 & 768.69 & 2383.20 & 14.49 & V:MT,  \\
1dd0e7ca1c & Na & \ce{Na3MnCr3O8} & 166 (R$\overline{3}$m) & 3.23 & 636.22 & 2551.29 & 13.20 & 3.29 & 649.36 & 2604.01 & 17.71 & Cr:MT,  \\
3668533566 & Na & \ce{NaVFeO4} & 62 (Pnma) & 2.38 & 329.80 & 1198.43 & 19.70 & 2.85 & 393.65 & 1430.44 & 6.98 & V:MT,  \\
0ac3187dad & Na & \ce{NaCrPO4} & 62 (Pnma) & 2.32 & 365.65 & 1248.16 & 21.48 & 3.07 & 484.50 & 1653.85 & 16.17 & Cr:MT,  \\
79b2af4a5f & K & \ce{K6MnFeF12} & 2 (P$\overline{1}$) & 5.41 & 1517.67 & 4142.30 & 106.38 & 3.43 & 961.53 & 2624.39 & 23.86 & F:MT,  \\
27c9f7324a & K & \ce{K6VFeF12} & 2 (P$\overline{1}$) & 5.16 & 1456.72 & 3948.23 & 38.50 & 3.50 & 989.34 & 2681.45 & 21.41 & F:MT, V:MT,  \\
faa07baa45 & K & \ce{K4CaFe3F14} & 12 (C2/m) & 5.05 & 859.12 & 2525.44 & 11.38 & 3.52 & 599.10 & 1761.11 & 16.77 & F:MT,  \\
baf4d3d5bd & K & \ce{K4CaMnFe2F14} & 12 (C2/m) & 4.85 & 826.92 & 2393.50 & 38.80 & 3.49 & 594.53 & 1720.87 & 18.67 & F:MT,  \\
45a32d2d03 & K & \ce{K4MnFe3F14} & 12 (C2/m) & 4.74 & 787.27 & 2456.59 & 32.86 & 3.63 & 602.99 & 1881.55 & 19.88 & F:MT,  \\
78cb10cd79 & K & \ce{K4CaMnCr2F14} & 12 (C2/m) & 4.66 & 803.33 & 2325.65 & 39.64 & 4.13 & 712.47 & 2062.60 & 21.35 & Cr:MT, F:MT,  \\
bb0f344d09 & K & \ce{KMn3CoO8} & 12 (C2/m) & 4.39 & 301.13 & 1066.49 & 7.91 & 2.34 & 160.72 & 569.22 & 14.34 & Co:MC,  \\
ab6e151f05 & K & \ce{K2VF5} & 12 (C2/m) & 4.39 & 1049.20 & 3019.92 & 9.23 & 3.79 & 906.98 & 2610.57 & 22.04 & F:MT, V:MT,  \\
dd00bcc6e3 & K & \ce{K2V(SiO3)3} & 11 (P2$_{1}$/m) & 4.28 & 641.77 & 1864.56 & 6.38 & 1.84 & 275.29 & 799.80 & 5.78 & V:MT,  \\
75ae834ca3 & K & \ce{K2V(Si2O5)3} & 2 (P$\overline{1}$) & 4.21 & 419.67 & 1138.11 & 5.77 & 1.82 & 181.37 & 491.86 & 3.61 & V:MT,  \\
efcb76889f & K & \ce{K4CaMnV2F14} & 12 (C2/m) & 3.99 & 691.32 & 1964.76 & 40.65 & 3.48 & 602.03 & 1710.99 & 18.88 & F:MT, V:MT,  \\
05b7811e98 & K & \ce{K4Fe3CoP4(O7F2)2} & 14 (P2$_{1}$/c) & 3.77 & 500.68 & 1599.76 & 18.64 & 3.99 & 530.07 & 1693.68 & 8.07 & F:MT, Co:MC,  \\
a7483104f0 & K & \ce{K8MgFe7P8(O7F2)4} & 2 (P$\overline{1}$) & 3.74 & 508.77 & 1587.62 & 14.90 & 3.89 & 528.92 & 1650.49 & 7.56 & F:MT,  \\
016d5b7e2b & K & \ce{K4CaV3F14} & 12 (C2/m) & 3.72 & 648.47 & 1846.75 & 36.73 & 3.48 & 605.67 & 1724.86 & 17.05 & F:MT, V:MT,  \\
faf804fc94 & K & \ce{K8CaFe7P8(O7F2)4} & 2 (P$\overline{1}$) & 3.70 & 498.87 & 1572.30 & 12.31 & 3.90 & 525.28 & 1655.57 & 7.56 & F:MT,  \\
494ad81024 & K & \ce{K4Mn3VF12} & 2 (P$\overline{1}$) & 3.40 & 606.72 & 1974.55 & 31.94 & 3.40 & 607.66 & 1977.60 & 30.96 & F:MT, V:MT,  \\
fb59e7ec3f & K & \ce{K4MnV3F12} & 2 (P$\overline{1}$) & 2.75 & 497.92 & 1631.62 & 49.87 & 3.45 & 625.23 & 2048.83 & 24.71 & F:MT, V:MT,  \\
d78228b1e2 & K & \ce{K(VS)2} & 139 (I4/mmm) & 2.13 & 278.06 & 1010.27 & 16.49 & 1.62 & 211.18 & 767.26 & 8.51 & V:MT,  \\
676610c71b & Mg & \ce{MgV2F12} & 2 (P$\overline{1}$) & 5.59 & 846.28 & 2296.60 & 26.37 & 3.38 & 510.94 & 1386.58 & 3.47 & V:MT, F:MT,  \\
1127401068 & Mg & \ce{SrMgVO4F} & 15 (C2/c) & 4.33 & 944.53 & 3501.38 & 9.22 & 3.39 & 739.28 & 2740.51 & 1.86 & F:MT, V:MT,  \\
5f9c619341 & Mg & \ce{Sr12Mg5Ni(PO4)12} & 2 (P$\overline{1}$) & 4.20 & 475.01 & 1720.98 & 1.32 & 1.98 & 224.20 & 812.28 & 3.32 & Ni:MC,  \\
7917258f43 & Mg & \ce{SrMgVP2O9} & 62 (Pnma) & 3.96 & 575.93 & 1935.82 & 7.80 & 3.26 & 473.75 & 1592.37 & 2.20 & V:MT,  \\
49f985fcbb & Mg & \ce{CaMgVO4F} & 15 (C2/c) & 3.87 & 1046.32 & 3334.98 & 2.00 & 3.26 & 881.01 & 2808.08 & 1.49 & F:MT, V:MT,  \\
6de10f712b & Ca & \ce{Ca3Ni(PO3)8} & 2 (P$\overline{1}$) & 5.59 & 1108.14 & 2958.21 & 23.02 & 5.31 & 1052.61 & 2809.97 & 8.37 & Ni:MC,  \\
c031c08efc & Ca & \ce{Ca3Co(PO3)8} & 2 (P$\overline{1}$) & 5.43 & 1077.03 & 2854.07 & 7.09 & 5.21 & 1033.79 & 2739.47 & 8.61 & Co:MC,  \\
6e82276b80 & Ca & \ce{Ca2CrP3O11} & 14 (P2$_{1}$/c) & 5.12 & 1368.73 & 4356.99 & 14.64 & 4.46 & 1192.69 & 3796.61 & 2.50 & Cr:MT,  \\
e014a307c2 & Ca & \ce{Ca2Ni(PO4)2} & 14 (P2$_{1}$/c) & 5.06 & 1650.67 & 5064.19 & 3.98 & 4.46 & 1455.36 & 4464.99 & 6.18 & Ni:MC,  \\
c5496e6797 & Ca & \ce{Ca2Co(PO4)2} & 14 (P2$_{1}$/c) & 4.81 & 1568.50 & 4815.61 & 0.00 & 4.31 & 1402.91 & 4307.20 & 5.71 & Co:MC,  \\
bb2bc39846 & Ca & \ce{CaFe2Ni(P2O7)2} & 14 (P2$_{1}$/c) & 3.48 & 334.49 & 1200.08 & 6.95 & 4.52 & 433.79 & 1556.35 & 1.64 & Ni:MC,  \\
\hline
\end{tabular}
\end{sidewaystable}

\begin{sidewaystable}
\centering
\caption{Candidates from the MACE screening with at least one element of unfeasible cost following Ref. \cite{BORAH2020OnBattery} classification, where MC: marginal cost, MT: marginal toxicity and UC: Unfeasible cost.}\label{Table_3}
\begin{tabular}{llll|cccc|cccc|m{4cm}}
\hline
& & & & \multicolumn{4}{|c|}{MACE-r2scan} & \multicolumn{4}{c|} {Energy-GNoME prediction} &  \\
 \cline{5-12}
GNOME id & ion & cathode & space group & V$_{\mathrm{avg}}$  & E$_{\mathrm{grav}}$ & E$_{\mathrm{vol}}$ & $\Delta$ V & V$_{\mathrm{avg}}$ & E$_{\mathrm{grav}}$ & E$_{\mathrm{vol}}$ & $\Delta$ V  & Elements \\
 &  & & &(V) & (Wh/kg) & (Wh/L) & \% & (V) & (Wh/kg) & (Wh/L) & \% & feasibility \\
\hline
e57f604a32 & Li & \ce{Li2P4RuO13} & 2 (P$\overline{1}$) & 5.56 & 667.54 & 1829.24 & 9.52 & 4.25 & 509.92 & 1397.33 & 1.99 & Li:MC, Ru:UC,  \\
9800894215 & Li & \ce{Li2CrP3WO12} & 33 (Pna2$_{1}$) & 3.02 & 303.26 & 1140.14 & 8.37 & 3.55 & 356.17 & 1339.10 & 3.60 & Cr:MT, Li:MC, W:UC,  \\
92ec723936 & Na & \ce{Na2Pd(CO3)2} & 14 (P2$_{1}$/c) & 4.02 & 790.21 & 2554.10 & 54.13 & 2.50 & 492.53 & 1591.94 & 1.91 & Pd:UC,  \\
6007a969e0 & Na & \ce{Na3Cr3GeO8} & 166 (R$\overline{3}$m) & 3.30 & 624.34 & 2571.59 & 7.62 & 3.38 & 638.34 & 2629.25 & 13.51 & Ge:UC, Cr:MT,  \\
0e4835cba8 & Na & \ce{Na6(RhO2)7} & 148 (R$\overline{3}$) & 3.30 & 489.85 & 2785.01 & 30.51 & 2.02 & 300.17 & 1706.64 & 7.93 & Rh:UC,  \\
e322f8fe63 & Na & \ce{Na3Mo(PO4)2} & 12 (C2/m) & 3.12 & 706.35 & 2416.41 & 15.79 & 2.87 & 649.83 & 2223.07 & 16.94 & Mo:UC,  \\
ee222a6d0d & Na & \ce{Na3P2W3O14} & 14 (P2$_{1}$/c) & 3.11 & 275.50 & 1479.72 & 4.39 & 2.80 & 247.94 & 1331.73 & 3.69 & W:UC,  \\
dafecaf0d0 & Na & \ce{NaMo(CO3)2} & 148 (R$\overline{3}$) & 3.10 & 348.23 & 1194.28 & 2.99 & 2.57 & 288.20 & 988.41 & 0.73 & Mo:UC,  \\
e772967865 & Na & \ce{Na3Ni2IrO6} & 2 (P$\overline{1}$) & 3.05 & 517.48 & 3103.21 & 18.40 & 2.70 & 457.24 & 2741.98 & 9.00 & Ir:UC, Ni:MC,  \\
5b56d416dd & Na & \ce{Na4Ni3IrO8} & 166 (R$\overline{3}$m) & 3.03 & 551.31 & 3095.43 & 20.11 & 2.78 & 506.33 & 2842.84 & 8.89 & Ir:UC, Ni:MC,  \\
945e0f6fc8 & Na & \ce{Na3SiMoCO7} & 11 (P2$_{1}$/m) & 2.65 & 671.71 & 2112.38 & 13.96 & 2.69 & 681.70 & 2143.80 & 2.19 & Mo:UC,  \\
11f803439e & K & \ce{K4CaMnGa2F14} & 12 (C2/m) & 5.63 & 918.14 & 2843.68 & 51.01 & 3.43 & 559.63 & 1733.28 & 19.57 & F:MT, Ga:UC,  \\
8c7aab7488 & K & \ce{KBe2P3O10} & 15 (C2/c) & 5.26 & 454.37 & 1083.54 & 7.73 & 4.05 & 350.39 & 835.59 & 3.32 & Be:UC,  \\
3833e50f4a & K & \ce{KSc(CO3)2} & 148 (R$\overline{3}$) & 5.13 & 674.09 & 1712.07 & 9.45 & 2.37 & 311.41 & 790.94 & 5.30 & Sc:UC,  \\
21ed59a268 & K & \ce{K3LiIr2O9} & 63 (Cmcm) & 4.90 & 603.65 & 3244.07 & 8.22 & 2.10 & 258.16 & 1387.39 & 9.24 & Ir:UC, Li:MC,  \\
1623782115 & K & \ce{K4BaTa5O15} & 12 (C2/m) & 3.94 & 293.47 & 2108.34 & 6.60 & 1.71 & 127.36 & 914.95 & 9.02 & Ta:UC,  \\
445080c82b & K & \ce{K4SrTa5O15} & 12 (C2/m) & 3.86 & 297.75 & 2087.24 & 6.82 & 1.72 & 132.62 & 929.71 & 9.07 & Ta:UC,  \\
4873aa021e & K & \ce{K2BaNb3O9} & 164 (P$\overline{3}$m1) & 3.47 & 291.38 & 1503.72 & 4.86 & 1.58 & 132.41 & 683.35 & 7.00 & Nb:UC,  \\
862c4a7a58 & K & \ce{K2SrNb3O9} & 164 (P$\overline{3}$m1) & 3.40 & 309.55 & 1501.80 & 4.80 & 1.59 & 144.80 & 702.49 & 7.03 & Nb:UC,  \\
061c0a3d69 & Mg & \ce{MgTa2F12} & 2 (P$\overline{1}$) & 5.96 & 520.16 & 2218.82 & 417.69 & 1.17 & 102.33 & 436.50 & 3.70 & Ta:UC, F:MT,  \\
b2d8ac9590 & Mg & \ce{MgZn2Rh3O8} & 166 (R$\overline{3}$m) & 5.87 & 531.53 & 3295.54 & 8.37 & 0.78 & 70.33 & 436.03 & 1.81 & Zn:MC, Rh:UC,  \\
250582d74b & Mg & \ce{MgNb2F12} & 2 (P$\overline{1}$) & 5.74 & 702.06 & 2277.94 & 20.20 & 2.83 & 345.94 & 1122.47 & 5.35 & Nb:UC, F:MT,  \\
58a98599bc & Mg & \ce{MgRu2F12} & 2 (P$\overline{1}$) & 4.65 & 548.37 & 1804.34 & 450.29 & 1.95 & 229.65 & 755.62 & 3.77 & Ru:UC, F:MT,  \\
d5232ca7a0 & Mg & \ce{MgMn2(AgO2)3} & 12 (C2/m) & 4.64 & 448.67 & 2707.45 & 14.06 & 3.14 & 304.23 & 1835.83 & 8.06 & Ag:UC,  \\
124aa7a44f & Mg & \ce{MgIr3F12} & 166 (R$\overline{3}$m) & 4.49 & 290.12 & 1384.02 & 1.40 & 8.39 & 542.51 & 2588.01 & 4.48 & Ir:UC, F:MT,  \\
b1cd5e0230 & Mg & \ce{Mg2CrGaAg4F14} & 12 (C2/m) & 4.31 & 532.60 & 2535.57 & 12.84 & 1.39 & 171.63 & 817.09 & 3.69 & Cr:MT, Ag:UC, F:MT, Ga:UC,  \\
7e9e66b9cc & Mg & \ce{MgRu3F12} & 2 (P$\overline{1}$) & 4.19 & 404.16 & 1732.24 & 12.61 & 1.78 & 171.68 & 735.81 & 3.74 & Ru:UC, F:MT,  \\
396537f2e5 & Mg & \ce{Mg2ScBi(Pt2O7)2} & 12 (C2/m) & 4.01 & 329.00 & 2822.21 & 3.12 & 3.68 & 302.11 & 2591.60 & 7.64 & Bi:UC, Sc:UC, Pt:UC,  \\
d0239de2c7 & Mg & \ce{MgOs2F12} & 2 (P$\overline{1}$) & 3.86 & 327.02 & 1467.68 & 1.64 & 2.46 & 208.20 & 934.41 & 3.69 & Os:UC, F:MT,  \\
342ada0410 & Mg & \ce{Ca2MgIrO6} & 14 (P2$_{1}$/c) & 3.79 & 516.97 & 2833.14 & 4.30 & 3.77 & 514.94 & 2821.99 & 4.72 & Ir:UC,  \\
635ee11f89 & Mg & \ce{Gd2MgMnO6} & 14 (P2$_{1}$/c) & 3.76 & 411.41 & 2908.26 & 4.79 & 3.26 & 356.97 & 2523.38 & 2.44 & Gd:UC,  \\
2f54381a7b & Mg & \ce{Sm4Mg2MnCoO12} & 2 (P$\overline{1}$) & 3.71 & 415.91 & 2871.59 & 4.91 & 1.34 & 150.33 & 1037.92 & 9.88 & Co:MC, Sm:UC,  \\
2e8aebdbc1 & Mg & \ce{Nd4Mg2MnCoO12} & 2 (P$\overline{1}$) & 3.71 & 426.48 & 2803.44 & 5.88 & 1.62 & 186.10 & 1223.32 & 1.37 & Nd:UC, Co:MC,  \\

\hline
\end{tabular}
\end{sidewaystable}

\setcounter{table}{2}
\begin{sidewaystable}
\centering
\caption{Continue: Candidates from the MACE screening with at least one element of unfeasible cost following Ref. \cite{BORAH2020OnBattery} classification, where MC: marginal cost, MT: marginal toxicity and UC: Unfeasible cost.}\label{Table_3}
\begin{tabular}{llll|cccc|cccc|l}
\hline
& & & & \multicolumn{4}{|c|}{MACE-r2scan} & \multicolumn{4}{c|} {Energy-GNoME prediction} &  \\
 \cline{5-12}
GNOME id & ion & cathode & space group & V$_{\mathrm{avg}}$  & E$_{\mathrm{grav}}$ & E$_{\mathrm{vol}}$ & $\Delta$ V & V$_{\mathrm{avg}}$ & E$_{\mathrm{grav}}$ & E$_{\mathrm{vol}}$ & $\Delta$ V  & Elements \\
 &  & & &(V) & (Wh/kg) & (Wh/L) & \% & (V) & (Wh/kg) & (Wh/L) & \% & feasibility \\
\hline
bd64810b56 & Mg & \ce{Nd2MgCoO6} & 14 (P2$_{1}$/c) & 3.68 & 421.94 & 2842.06 & 4.84 & 1.68 & 192.12 & 1294.04 & 1.40 & Nd:UC, Co:MC,  \\
29cfae4cf5 & Mg & \ce{Nd2MgPtO6} & 14 (P2$_{1}$/c) & 3.51 & 311.66 & 2488.85 & 4.98 & 0.59 & 52.08 & 415.91 & 2.88 & Nd:UC, Pt:UC,  \\
476290b0a7 & Mg & \ce{Tb2MgMnO6} & 14 (P2$_{1}$/c) & 3.27 & 355.52 & 2574.26 & 8.05 & 3.21 & 349.09 & 2527.69 & 2.17 & Tb:UC,  \\
2a864e4e72 & Mg & \ce{MgMn2Rh3O8} & 166 (R$\overline{3}$m) & 3.16 & 297.05 & 1709.32 & 11.87 & 1.81 & 169.86 & 977.41 & 2.80 & Rh:UC,  \\
4a7b5ac649 & Mg & \ce{MgNiPd3O8} & 166 (R$\overline{3}$m) & 3.08 & 311.41 & 1712.84 & 11.65 & 1.67 & 168.65 & 927.61 & 3.36 & Pd:UC, Ni:MC,  \\
2241f29998 & Mg & \ce{MgReF6} & 148 (R$\overline{3}$) & 3.08 & 508.07 & 2378.73 & 145.26 & 4.20 & 693.64 & 3247.56 & 1.21 & Re:UC, F:MT,  \\
b4b2812e0a & Mg & \ce{MgNi3(PdO4)2} & 2 (P$\overline{1}$) & 2.83 & 280.64 & 1623.22 & 6.68 & 1.97 & 195.34 & 1129.87 & 5.38 & Pd:UC, Ni:MC,  \\
70289d5f4a & Mg & \ce{YbMg(SiO3)2} & 15 (C2/c) & 2.76 & 423.92 & 2205.08 & 13.27 & 3.10 & 475.39 & 2472.80 & 4.46 & Yb:UC,  \\
f9dc9dd90e & Mg & \ce{MgV2TeO7} & 2 (P$\overline{1}$) & 2.73 & 400.47 & 1646.82 & 0.00 & 2.30 & 336.34 & 1383.08 & 7.76 & V:MT, Te:UC,  \\
7eca1d34a5 & Ca & \ce{Ca(Pd3O4)2} & 225 (Fm$\overline{3}$m) & 5.22 & 347.16 & 2040.88 & 9.78 & 2.18 & 144.67 & 850.45 & 6.66 & Pd:UC,  \\
e77ca2d69a & Ca & \ce{Ca2Rh3O8} & 12 (C2/m) & 4.33 & 898.95 & 4519.17 & 21.11 & 2.86 & 592.38 & 2978.03 & 7.36 & Rh:UC,  \\
9abb7a5783 & Ca & \ce{CaAg3(RhO3)2} & 12 (C2/m) & 3.52 & 283.22 & 1858.50 & 14.95 & 2.57 & 206.61 & 1355.80 & 7.63 & Ag:UC, Rh:UC,  \\

\hline
\end{tabular}
\end{sidewaystable}
\clearpage

\end{document}